\documentclass[aps,prd,twocolumn,groupedaddress]{revtex4-1}

\newcommand{\ds}{\displaystyle}
\usepackage{enumerate,graphicx}

\begin{document}


\title{Scalar field excited around a rapidly rotating black hole
       in Chern-Simons modified gravity }


\author{Kohkichi Konno}
\email[]{kohkichi@gt.tomakomai-ct.ac.jp}
\affiliation{Department of Natural and Physical Sciences,
  Tomakomai National College of Technology, 
  Tomakomai 059-1275, Japan}

\author{Rohta Takahashi}
\affiliation{Department of Natural and Physical Sciences,
  Tomakomai National College of Technology, 
  Tomakomai 059-1275, Japan}


\date{\today}

\begin{abstract}
We discuss a Chern-Simons (CS) scalar field around a rapidly 
rotating black hole in dynamical CS modified gravity. 
The CS correction can be obtained perturbatively by considering 
the Kerr spacetime to be the background. We obtain the CS scalar 
field solution around the black hole analytically and numerically, 
assuming a stationary and axisymmetric configuration.
The scalar field diverges on the inner horizon
when we impose the boundary condition that the scalar field 
is regular on the outer horizon and vanishes at infinity.
Therefore, the CS scalar field becomes problematic 
on the inner horizon.
\end{abstract}


\maketitle

\section{Introduction}

The ultimate theory for gravity is an open question. 
General relativity has been repeatedly tested 
in the weak gravitational field \cite{will}. 
However, the properties of the strong gravitational 
field have not yet been fully verified. As for modifications 
in the strong gravitational field, the possibilities 
of various gravity theories still remain.

One of the most promising alternative gravity theories is
dynamical Chern-Simons (CS) modified gravity 
(see e.g.~\cite{ay_review}).
This theory can be derived both from an effective theory
of superstring theory \cite{seck} and 
from that of loop quantum gravity \cite{ty,mercuri}. 
Furthermore, the CS modified gravity is derived from
an effective field theory for inflation \cite{weinberg}.
Such a theory contains corrections consisting of 
higher-order derivative terms. Thus high curvature,
namely, strong field regime such as the vicinity of a 
black hole is inevitably modified.
The CS modified gravity was originally formulated
in a non-dynamical theory \cite{jp}, in which a presumed, 
non-dynamical scalar field is coupled to the gravitational field.
However, the non-dynamical theory has several problems: 
(i) curvature singularities appear on the rotational axis
for a rotating black hole \cite{kmt1,kmat}
(see also \cite{gy} for the comprehensive study of a rotating 
black hole), 
(ii) a wide class of black hole oscillation modes 
is suppressed \cite{ys}, and 
(iii) ghost appears in general \cite{ms}.
Hence many authors 
\cite{yp,kmt2,sy,cg,hkl,ypol,mpcg,
aeg,gpy,yooa,sty,pcg,ac,ms,yyt1,dfk,yyt2,ysyt,st,vincent}
have come to discuss the dynamical CS modified gravity, 
in which the kinetic term for the scalar field 
is taken into account, to pursue self-consistency for the theory.
In the dynamical theory, the first two problems mentioned above 
do not occur \cite{yp,kmt2,yyt1,cg,mpcg,gpy,pcg}, 
and the last problem does not also appear at least 
in a certain situation \cite{cg,mpcg,ms,dfk}. 
Therefore, the dynamical CS modified gravity 
has attract much attention in recent years.

The most remarkable feature of the dynamical CS modified gravity
is that the Schwarzschild spacetime is the unique solution 
for a static and asymptotically flat black hole \cite{st}
(see also \cite{roga} for the case of a charged black hole).
Similarly, the spacetime outside a static and spherically symmetric star
becomes the Schwarzschild spacetime (see e.g. \cite{ac}) 
when we assume asymptotically flat spacetime again, 
because CS corrections exactly vanishes in that case. 
Thus the dynamical CS modified gravity gives the same 
result as that in general relativity 
for the leading term at the post-Newtonian order.  
Hence, this theory passes the classical tests for general relativity. 
Once the assumption of a static state 
is broken, spacetimes in the dynamical CS modified gravity
could deviate from those in general relativity. 
As an example for such spacetimes, we can consider 
a stationary spacetime around a rotating black hole.
The deviation from general relativity becomes large when we approach 
strong field regime such as the vicinity of a rotating black hole.
In weak field regime, we could merely derive small deviation 
from general relativity. In fact, the experimental results of 
Gravity Probe B and the LAGEOS satellites around the earth
give a rather weak constraint for a CS coupling constant $\alpha$ 
such that $\sqrt{\alpha} \lesssim 10^{8} [{\rm km}]$ \cite{ac}
(see Sec.~\ref{sec:csg} for the definition of the coupling 
constant). Nevertheless, this is the most stringent constraint 
ever obtained. Therefore, we need to investigate the strong field 
regime without assuming a static state
to test the validity of the dynamical CS modified gravity.

In this paper, we consider a stationary spacetime around 
a rapidly rotating black hole. In the previous works 
\cite{yp,kmt2,yyt1}, both slow rotation and weak CS 
coupling are assumed for discussing rotating black holes. 
However, the former assumption is insufficient for extracting 
distinctive features of the dynamical CS modified gravity 
thoroughly. Thus we do not rely on the slow rotation 
approximation in the present work. As mentioned above, 
we can regard the dynamical CS modified gravity as 
an effective theory of more fundamental theories.
Thus, the assumption of weak CS coupling would 
still be valid, and we can treat CS corrections in the 
framework of the perturbation theory. To deal with a rapidly 
rotating black hole, we adopt the Kerr spacetime as the background.  
As explained below, the CS scalar field and the CS 
gravitational correction are determined alternately in a 
bootstrapping scheme. Therefore, we investigate the CS scalar 
field around a rapidly rotating black hole as the first step 
for the CS corrections to the stationary spacetime.

This paper is organized as follows. In Sec.~\ref{sec:csg},
we briefly review the framework of the dynamical 
CS modified gravity. In Sec.~\ref{sec:CS-c}, 
we discuss CS corrections to a rapidly rotating black hole
based on the perturbation theory. We obtain the 
formal solution for the CS scalar field around 
the black hole, assuming a stationary and axisymmetric 
configuration. In particular, we obtain the analytic solution
for the leading term in the Legendre expansion of the 
scalar field with respect to the polar angle. 
In Sec.~\ref{sec:prop-sf}, we discuss the properties 
of the scalar field solution that can be obtained by 
numerical integration. Finally, we provide a conclusion 
in Sec.~\ref{sec:concl}. We use the geometrical units 
in which $c=G=1$ throughout the paper and follow the 
conventions of \cite{weinberg2}.

\section{Dynamical CS modified gravity}
\label{sec:csg}

The action of the dynamical CS modified gravity is provided by 
\cite{jp,ay_review}
\begin{eqnarray}
 S & = & \int d^{4}x \sqrt{-g} \left[ - \kappa R + 
   \frac{\kappa\alpha}{4} \vartheta 
   \: ^{\ast}\! R^{\tau\sigma\mu\nu} R_{\sigma\tau\mu\nu} \right.
   \nonumber \\
 & & \left. 
 - \frac{1}{2} g^{\mu\nu} \left( \partial_{\mu} \vartheta \right)
   \left( \partial_{\nu} \vartheta \right) 
   -  V\left( \vartheta \right) + {\cal L}_{\rm m} \right] ,
\label{eq:action}
\end{eqnarray}
where $g$ is the determinant of the metric $g_{\mu\nu}$, 
$R:= g^{\mu\nu} R_{\mu\nu} $ is the 
Ricci scalar defined with the Ricci tensor
$R_{\mu\nu}:=R^{\lambda}_{\ \mu\lambda\nu} $, 
$R^{\mu}_{\ \nu\lambda\sigma} 
 := \partial_{\sigma} \Gamma^{\mu}_{\nu\lambda}-\cdots$ 
is the Riemann tensor, and
$ ^{\ast}\! R^{\tau\sigma\mu\nu} := \frac{1}{2} 
 \varepsilon^{\mu\nu\alpha\beta} R^{\tau\sigma}_{\quad \alpha\beta}$
is the dual Riemann tensor. Here, $\Gamma^{\mu}_{\nu\lambda}$ 
stands for the Christoffel symbols, and 
$\varepsilon^{\mu\nu\alpha\beta} $ is the Levi-Civita tensor
with the definition of $\varepsilon^{0123} := 1/\sqrt{-g}$. 
In the action, $\vartheta$ is a scalar field with 
the dimension of $\left[ \vartheta \right] = L^{0}$, 
$V\left( \vartheta \right)$ is the potential for $\vartheta$, 
and ${\cal L}_{\rm m}$ is the Lagrangian for matter.
The coefficient $\kappa :=\left( 16\pi \right)^{-1}$ has the 
dimension of $\left[ \kappa \right]=L^{0}$, 
while the coupling constant $\alpha$ has the dimension
of $\left[ \alpha \right] = L^{2}$.
The first term in Eq.~(\ref{eq:action}) is the Einstein-Hilbert 
action. The second term, which is called the Chern-Pontryagin 
(CP) term, can be converted to the CS term 
via partial integration as 
\begin{equation}
 -\frac{\kappa\alpha}{2} \int d^4 x \sqrt{-g} 
 \varepsilon^{\mu\nu\sigma\lambda} \partial_{\mu} \vartheta 
\left( \Gamma^{\alpha}_{\nu \beta} \partial_{\sigma} 
  \Gamma^{\beta}_{\lambda\alpha} + \frac{2}{3} 
  \Gamma^{\gamma}_{\nu\alpha} 
  \Gamma^{\alpha}_{\sigma\beta} \Gamma^{\beta}_{\lambda\gamma} 
  \right) .
\end{equation}
The third term in Eq.~(\ref{eq:action}) is the kinematic term 
for $\vartheta$. Throughout the paper, we neglect the potential 
$V(\vartheta )$ for simplicity (see \cite{ay_review} 
for justification in relation to the string theory).
While the kinematic term for $\vartheta$ 
is neglected in the nondynamical theory \cite{jp}, 
it is taken into account in the dynamical theory.

 From the variations with respect to the metric $g_{\mu\nu}$ 
and the scalar field $\vartheta$, we derive 
the field equations in the forms 
\begin{equation}
\label{eq:feq-g}
 G^{\mu\nu} + \alpha C^{\mu\nu} 
 = - \frac{1}{2\kappa} \left( 
   T_{\rm m}^{\mu\nu} + T_{\vartheta}^{\mu\nu} \right) ,
\end{equation}
\begin{equation}
\label{eq:feq-s}
 g^{\mu\nu} \nabla_{\mu} \nabla_{\nu} \vartheta
 = - \frac{\kappa \alpha}{4} \:
  ^{\ast} R^{\tau\sigma\mu\nu} R_{\sigma\tau\mu\nu} ,
\end{equation}
where $G^{\mu\nu} := R^{\mu\nu} - \frac{1}{2} g^{\mu\nu}R$
is the Einstein tensor, and $C^{\mu\nu}$ is called $C$ tensor 
\cite{ay_review} and defined as
\begin{equation}
 C^{\mu\nu} 
 := \left( \nabla_{\alpha} \vartheta \right) 
     \varepsilon^{\alpha\beta\gamma (\mu} \nabla_{\gamma} 
     R^{\nu)}_{\ \beta} + 
   \left( \nabla_{\sigma} \nabla_{\lambda} \vartheta \right)
   \: ^{\ast}\! R^{\lambda ( \mu\nu ) \sigma}. 
\end{equation}
Here, round brackets for indexes denote symmetrization such as
$A^{(\mu\nu)} = \frac{1}{2} \left( A^{\mu\nu}+A^{\nu\mu}\right)$.
In the field equation (\ref{eq:feq-g}), $T_{\rm m}^{\mu\nu}$ is 
the stress-energy tensor for usual matter, and 
$T_{\vartheta}^{\mu\nu}$ is that for the scalar field 
$\vartheta$ as
\begin{equation}
 T_{\vartheta}^{\mu\nu} = \left( \nabla^{\mu} \vartheta \right)
  \left( \nabla^{\nu} \vartheta \right)
-\frac{1}{2} g^{\mu\nu} \left( \nabla^{\lambda} \vartheta \right)
  \left( \nabla_{\lambda} \vartheta \right) .
\end{equation}
In this paper, we consider vacuum for usual matter, 
i.e., $T_{\rm m}^{\mu\nu} = 0$. 
Thus Eqs.~(\ref{eq:feq-g}) and (\ref{eq:feq-s}) govern
the dynamics of the gravitational field and the scalar field
in the dynamical CS modified gravity.

\section{CS corrections to a rapidly rotating black hole}
\label{sec:CS-c}

\subsection{Outline of the perturbative approach}

The CS modified gravity is regarded as an effective 
theory obtained from more fundamental theories.
Thus the CS corrections are considered to be 
small in comparison with the terms appearing 
in general relativity. The CS corrections 
can be tuned with the coupling constant $\alpha$. 
Hence we can assume $\alpha \ll {\cal M}^2$, 
where ${\cal M}$ denotes a mass scale in question,
in particular, it corresponds to a black hole mass in our problem.

We can treat the CS corrections in the framework of the 
perturbation theory with the expansion parameter
$\tilde{\alpha} := \alpha / {\cal M}^2$. 
We assume that the background solution $g^{(0)}_{\mu\nu}$
is a solution in general relativity and that the scalar field
completely vanishes in the background, i.e., $\vartheta^{(0)}=0$,
where $g^{(m)}_{\mu\nu}\sim O\left( \tilde{\alpha}^m \right)$ and 
$\vartheta^{(n)} \sim O\left( \tilde{\alpha}^n \right)$.
To expand the metric and the scalar field into power series 
in $\tilde{\alpha}$, let us discuss the structure of the basic 
equations (\ref{eq:feq-g}) and (\ref{eq:feq-s}).
Equation (\ref{eq:feq-g}) is now written as
\begin{equation}
 \label{eq:feq-g2}
 G^{\mu\nu} = - \alpha C^{\mu\nu} 
  - \frac{1}{2\kappa} T_{\vartheta}^{\mu\nu} .
\end{equation}
Equations (\ref{eq:feq-s}) and (\ref{eq:feq-g2}) can be
solved in a bootstrapping scheme. 
To understand the scheme, it may be useful to recall that 
the curvature tensors are essentially composed of 
the second-order derivatives of the metric and that 
$\partial_{\mu} \sim O\left( 1/ {\cal M}\right)$.  
The bootstrapping scheme is as follows.
First, given a background solution $g_{\mu\nu}=g^{(0)}_{\mu\nu}$, 
the righthand side of Eq.~(\ref{eq:feq-s}) provides
the first-order corrections in $\tilde{\alpha}$.
Thus the solution for $\vartheta^{(1)}$  
can be obtained from Eq.~(\ref{eq:feq-s}). 
Second, when we have $g_{\mu\nu}=g^{(0)}_{\mu\nu}$ 
and $\vartheta=\vartheta^{(1)}$, the righthand side of 
Eq.~(\ref{eq:feq-g2}) provides the second-order corrections
in $\tilde{\alpha}$. The solution for $g^{(2)}_{\mu\nu}$
can be obtained from Eq.~(\ref{eq:feq-g2}).
Next, given $g_{\mu\nu}=g^{(0)}_{\mu\nu}+g^{(2)}_{\mu\nu}$,
the solution for $\vartheta^{(3)}$ can be obtained 
from the terms of order $\tilde{\alpha}^3$ in Eq.~(\ref{eq:feq-s}). 
Then, given $g_{\mu\nu}=g^{(0)}_{\mu\nu}+g^{(2)}_{\mu\nu}$
and $\vartheta=\vartheta^{(1)}+ \vartheta^{(3)}$, 
the solution for $g^{(4)}_{\mu\nu}$ can be obtained 
from the terms of order $\tilde{\alpha}^4$ in Eq.~(\ref{eq:feq-g2}).
The same procedure is repeated to produce the power series 
in $\tilde{\alpha}$, 
\begin{eqnarray}
\label{eq:expand-th}
 \vartheta & = & \vartheta^{(1)} + \vartheta^{(3)} + \cdots , \\
\label{eq:expand-g}
 g_{\mu\nu} & = & g_{\mu\nu}^{(0)} + g_{\mu\nu}^{(2)} + \cdots . 
\end{eqnarray}
It should be noted that while $\vartheta$ is expanded by 
odd powers in $\tilde{\alpha}$, $g_{\mu\nu}$ is expanded by 
even powers in $\tilde{\alpha}$. 
Thus, we need to solve the scalar field $\vartheta$ and 
the gravitational field $g_{\mu\nu}$ alternately 
in the bootstrapping scheme. 
The series expansions in Eqs.~(\ref{eq:expand-th}) 
and (\ref{eq:expand-g}) is validated for the region
in which $\alpha$ is much smaller than the squared 
curvature radius. Thus these expansions would be valid
except for the neighborhood of the curvature singularity.

\subsection{Background metric}

To discuss a rapidly rotating black hole in the 
dynamical CS modified gravity, 
we adopt the Kerr spacetime as the background.
We start from the metric in the Boyer-Lindquist coordinates. 
After the coordinate transformation of $\mu=-\cos\theta$, 
where $\theta$ denotes the polar angle from the rotational axis, 
we derive the metric 
\begin{eqnarray}
\label{eq:kerr-m}
 ds^2 &=& g^{(0)}_{\mu\nu} dx^{\mu} dx^{\nu} 
 \nonumber \\
 & = & - \left( 1-\frac{2Mr}{\rho^2} \right) dt^2
 -\frac{4Mar}{\rho^2}\left( 1-\mu^2\right)  dtd\phi 
  \nonumber \\
 && +\frac{\Sigma}{\rho^2}\left( 1-\mu^2\right) d\phi^2 
 +\frac{\rho^2}{\Delta} dr^2 + \frac{\rho^2}{1-\mu^2} d\mu^2 , 
\end{eqnarray}
where $M$ is the mass of the black hole,  
$a:=J/M$ denotes the rotational parameter defined by 
the angular momentum $J$ of the black hole, and 
\begin{eqnarray}
\rho^2 & := & r^2 + a^2 \mu^2 , \\
 \Sigma & := & 
   \left( r^2 +a^2 \right) \rho^2 + 2Ma^2 r \left( 1-\mu^2\right) ,\\
 \Delta & := & r^2 -2Mr+a^2 .
\end{eqnarray}
We assume $0\leq a < M$ to avoid naked singularity.
 From the metric $g_{\mu\nu}^{(0)}$, 
we have $\sqrt{-g^{(0)}} = \rho^2$.
We can also calculate the CP term, which appears 
in the source term in Eq.~(\ref{eq:feq-s}), as
\begin{equation}
\label{eq:cp}
 ^{\ast} R^{\tau\sigma\nu\lambda} R_{\sigma\tau\nu\lambda} 
 = 96 M^2 \left( \frac{3ar\mu}{\rho^8} 
   - \frac{16a^3 r^3 \mu^3}{\rho^{12}}\right) .
\end{equation}
Thus the CP term becomes singular only at $\rho^2=0$, i.e., 
the curvature singularity of the Kerr spacetime.

\subsection{First-order CS scalar field}

The field equation (\ref{eq:feq-s}) for the scalar field 
$\vartheta^{(1)}$ can also be expressed as
\begin{equation}
\label{eq:feq-s2}
 \partial_{\nu} \left( \sqrt{-g} g^{\nu\lambda} 
   \partial_{\lambda} \vartheta^{(1)} \right)
 = - \frac{\kappa \alpha}{4} \sqrt{-g} \:
  ^{\ast} R^{\tau\sigma\nu\lambda} R_{\sigma\tau\nu\lambda} ,
\end{equation}
where $\vartheta$ depends on $r$ and $\mu$ due to
the assumption of stationarity and axisymmetry.
Using Eq.~(\ref{eq:cp}), we derive
\begin{eqnarray}
\label{eq:feq-s3}
 \frac{\partial}{\partial r} 
  \left[ \Delta \frac{\partial}{\partial r} 
   \vartheta^{(1)} (r,\mu )\right]
  + \frac{\partial}{\partial \mu} 
  \left[ \left( 1-\mu^2 \right) \frac{\partial}{\partial \mu} 
  \vartheta^{(1)} (r,\mu ) 
  \right]  \nonumber \\
 = - \frac{3}{2\pi} \alpha M^2 \left( \frac{3ar\mu}{\rho^6} 
   - \frac{16a^3 r^3 \mu^3}{\rho^{10}}\right) .
\end{eqnarray}
When we introduce dimensionless variables 
\begin{eqnarray}
 \tilde{r} & := & \frac{r}{M}, \\
 \tilde{a} & := & \frac{a}{M}, \\ 
 \tilde{\rho}^2 & := & \tilde{r}^2 + \tilde{a}^2 \mu^2 ,  
\end{eqnarray}
the field equation (\ref{eq:feq-s3}) becomes the form
\begin{eqnarray}
\label{eq:scalar-main}
 \frac{\partial}{\partial \tilde{r}} 
   \left[ \left( \tilde{r}^2 -2\tilde{r} + \tilde{a}^2\right) 
   \frac{\partial \vartheta^{(1)}}{\partial \tilde{r}} \right] 
   + \frac{\partial}{\partial \mu} 
   \left[ \left( 1-\mu^2 \right) 
   \frac{\partial \vartheta^{(1)}}{\partial \mu} \right] 
   \nonumber \\
 = \tilde{\alpha} S \left( \tilde{r}, \mu \right) , 
\end{eqnarray}
where $\tilde{\alpha} = \alpha / M^2$, and 
\begin{equation}
\label{eq:source-term}
 S \left( \tilde{r}, \mu \right) 
 := - \frac{3}{2\pi} \left( \frac{3\tilde{a}\tilde{r}\mu}{\tilde{\rho}^6} 
   - \frac{16\tilde{a}^3 \tilde{r}^3 \mu^3}{\tilde{\rho}^{10}} \right) .
\end{equation}
Equation (\ref{eq:scalar-main}) is the main equation 
for the scalar field $\vartheta^{(1)}$ 
of the first order in $\tilde{\alpha}$.

To solve Eq.~(\ref{eq:scalar-main}), 
we may expand the scalar field $\vartheta$ by 
the Legendre polynomials of the first kind $P_{n}(-\mu)$, 
because the set of $\left\{ P_{n} \left( -\mu \right)\right\}
=\left\{ P_{n} \left( \cos \theta \right)\right\}$ forms 
the system of orthogonal functions that are regular 
anywhere in the region of $0\leq \theta \leq \pi$. 
Then we have
\begin{equation}
\label{eq:theta-expand}
 \vartheta^{(1)} \left( \tilde{r}, \mu \right)
 = \tilde{\alpha} \sum_{n} \Theta_{n} 
   \left( \tilde{r}\right) P_{n} (-\mu ) ,
\end{equation}
where $n=0, 1, 2, \cdots$, and
$\Theta_{n}$ is a function of $\tilde{r}$ only.
We also expand the source term $S\left( \tilde{r}, \mu \right)$ as
\begin{equation}
\label{eq:s-expand}
 S\left( \tilde{r}, \mu \right) 
 = \sum_{n} S_{n} \left( \tilde{r}\right) P_{n}\left( -\mu \right) .
\end{equation}
Here the Legendre polynomials $P_{n}$ satisfies the differential 
equation
\begin{equation}
\label{eq:legendre-diff-eq}
 \frac{d}{d\mu} \left[ \left( 1-\mu^2 \right)
 \frac{d}{d\mu} P_{n} ( -\mu )\right]
 + n(n+1) P_{n} ( -\mu ) =0 .
\end{equation}
Substituting Eqs.~(\ref{eq:theta-expand}) and (\ref{eq:s-expand})
into Eq.~(\ref{eq:scalar-main}) and 
using Eq.~(\ref{eq:legendre-diff-eq}), we derive
the equation for the radial part, 
\begin{equation}
\label{eq:diff-eq-th-r}
 \frac{d}{d\tilde{r}} \left[ \left( \tilde{r}^2 -2\tilde{r} +a^2 \right) 
   \frac{d}{d\tilde{r}} \Theta_{m} \right]
   -m(m+1) \Theta_{m} = S_{m} , 
\end{equation}
where $m=0, 1,2, \cdots$, and we have used the formula
\begin{equation}
 \int_{-1}^{1} P_{n} \left( -\mu\right) P_{m} \left( -\mu \right) d\mu
 = \frac{2}{2n+1} \delta_{nm} .
\end{equation}
 From the functional form of Eq.~(\ref{eq:source-term}), we find
\begin{equation}
\label{eq:smr}
 S_{m} \left( \tilde{r}\right) = \left\{ 
 \begin{array}{lc}
  0 & \left(m: \mbox{even}\right) ,\\
 \ds \left( 2m+1 \right) \int_{0}^{1} S\left( \tilde{r},-\mu\right)
   P_{m} \left( \mu \right)d\mu
  & \left(m: \mbox{odd}\right) . 
 \end{array}\right. 
\end{equation}
For $m=0$, Eq.~(\ref{eq:diff-eq-th-r}) becomes
\begin{equation}
 \frac{d}{d\tilde{r}} \left[ \left( \tilde{r}^2 -2\tilde{r} +a^2 \right) 
   \frac{d}{d\tilde{r}} \Theta_{0} \right] = 0 , 
\end{equation}
This equation admits the solution $\Theta_{0} =\mbox{const}$, 
which corresponds to a shift symmetry 
$\vartheta \rightarrow \vartheta + \mbox{const}$.
For the remaining even part ($m=2n\geq2$), 
the source term in Eq.~(\ref{eq:diff-eq-th-r}) 
vanishes, and, therefore, we derive the trivial solution of 
$\Theta_{2n} =0$. Thus, we focus on the odd part ($m=2n+1$) 
only, hereafter. By using the formula
\begin{eqnarray}
  P_{2n+1} (\mu ) & = & \sum_{k=0}^{n} (-1)^k
  \frac{(4n+2-2k)!}{2^{2n+1} k!(2n+1-k)! (2n+1-2k)!} 
  \nonumber \\
 &&  \times \mu^{2n+1-2k} , 
\end{eqnarray}
we can obtain the odd modes $S_{2n+1}$ analytically as 
\begin{equation}
\label{eq:s-term}
 S_{2n+1} \left( \tilde{r} \right)
 = \sum_{k=1}^{2} \sum_{m=0}^{n} \sum_{l=0}^{m+k}
   c\left( n,m,k,l\right) I_{2m-1, l-2k} \left( \tilde{r} \right) , 
\end{equation}
where $c\left( n,m,k,l\right)$ is given by
\begin{eqnarray}
 c\left( n,m,k,l\right)
 & := & (-1)^{n+1-l} (4n+3) (k+2)^{k} 
  \frac{(m+k)!}{l!(m+k-l)!} 
  \nonumber \\
 &&  \times \frac{3\left[ 2\left(n+m+1\right)\right]!}
  {2^{2n+2} \pi\left( n-m\right)! \left( n+m+1\right)! (2m+1)!} ,
  \nonumber \\
 && 
\end{eqnarray}
and $I_{i,j}$ is defined as
\begin{enumerate}[(i)]
 \item for $j>0$,
       \begin{eqnarray}
	I_{i,j} \left( \tilde{r} \right) := 
         \sum_{p=0}^{j-1} B^{+}_{p} \left( i, j \right) 
         \tilde{r}^{i-2-2p} ,
       \end{eqnarray}
 \item for $j=0$,
       \begin{eqnarray}
	I_{i,0} \left( \tilde{r} \right) := 
         B^{0} \left( i, 0 \right) \tilde{r}^{i-1} 
         \arctan \frac{\tilde{a}}{\tilde{r}} ,
       \end{eqnarray}
 \item for $j<0$, 
       \begin{eqnarray}
	I_{i,j} \left( \tilde{r} \right) & := &
         \sum_{p=1}^{|j|} B^{-}_{p} \left( i,j \right)
         \frac{\tilde{r}^{i+2|j|-2p}}
         {\left( \tilde{r}^2+\tilde{a}^2 \right)^{|j|+1-p}}
         \nonumber \\
        && + B^{0} \left( i,j \right) \tilde{r}^{i-1} 
         \arctan \frac{\tilde{a}}{\tilde{r}}.
       \end{eqnarray}
\end{enumerate}
Here $B^{+}_{p}$, $B^{0}$ and $B^{-}_{p}$ are defined, 
respectively, as
\begin{eqnarray}
 B^{+}_{p} \left( i,j \right) 
  & := & \frac{(j-1)!}{p!(j-1-p)!}
    \frac{1}{2p+1} \frac{1}{\tilde{a}^{i+2-2p}} , \\
 B^{0} \left( i,j \right)
  & := & \frac{(2|j|)!}{2^{2|j|} \left(|j|!\right)^2} 
    \frac{1}{\tilde{a}^{i+3}} \\
 B^{-}_{p} \left( i,j \right) 
  & := & \frac{(2|j|)! (|j|+1-p)! (|j|-p)!}
    {2^{2p-1} \left[ 2\left( |j|+1-p\right)\right]! \left(|j|!\right)^2}
    \frac{1}{\tilde{a}^{i+2}} .
\end{eqnarray}
The explicit forms of $S_{2n+1}$ for lower orders 
are given in Appendix \ref{sec:S_forms}.
Thus the main equation (\ref{eq:scalar-main}) is reduced to 
the odd part ($m=2n+1$) of the ordinary differential 
equation (\ref{eq:diff-eq-th-r}) with the source term 
given by Eq.~(\ref{eq:s-term}).

To solve Eq.~(\ref{eq:diff-eq-th-r}), 
let us introduce the variable transformation
\begin{equation}
 \tilde{r} \rightarrow \tilde{\eta} 
  = \frac{\tilde{r}-1}{\sqrt{1-\tilde{a}^2}} .
\end{equation}
Note that the domain of $\tilde{r}\geq \tilde{r}_{+}$ is mapped 
onto the domain of $\tilde{\eta} \geq 1 $, where
$\tilde{r}_{+}:=1+\sqrt{1-\tilde{a}^2}$ corresponds to the radius of 
the outer horizon of the Kerr black hole. 
The differential equation for $\Theta_{2n+1}$ is 
then expressed as
\begin{eqnarray}
\label{eq:Th-diff-eq}
 \frac{1}{1-\tilde{\eta}^2} 
 \frac{d}{d\tilde{\eta}} \left[ \left( 1-\tilde{\eta}^2\right)
  \frac{d}{d\tilde{\eta}} \Theta_{2n+1} \right]
 + \frac{(2n+1)(2n+2)}{1-\tilde{\eta}^2} \Theta_{2n+1}
 \nonumber \\
 = - \frac{1}{1-\tilde{\eta}^2} S_{2n+1} 
   \left( \tilde{r} \left( \tilde{\eta}\right)\right) ,
 \nonumber \\
\end{eqnarray}
The independent solutions for the homogeneous equation 
of Eq.~(\ref{eq:Th-diff-eq}) are given by 
$P_{2n+1} \left(\tilde{\eta} \right)$
and $Q_{2n+1} \left(\tilde{\eta} \right)$, where $Q_{2n+1}$
is the Legendre polynomials of the second kind. 
The Wronskian ${\cal W}$ is then given by
\begin{eqnarray}
 {\cal W} \left( P_{2n+1} , Q_{2n+1} \right)& = & 
   P_{2n+1} \frac{dQ_{2n+1}}{d\tilde{\eta}}
    - \frac{dP_{2n+1}}{d\tilde{\eta}} Q_{2n+1}
    \nonumber \\ 
 & = & \frac{1}{1-\tilde{\eta}^2} .
\end{eqnarray}
Therefore the general solution of Eq.~(\ref{eq:Th-diff-eq}) 
is given by \cite{arfken} 
\begin{eqnarray}
\label{eq:g-s1}
 \Theta_{2n+1} (\tilde{\eta} )& = & 
  C_{2n+1}  P_{2n+1} \left(\tilde{\eta} \right) 
  + D_{2n+1} Q_{2n+1} \left(\tilde{\eta} \right) 
  \nonumber \\
 && + P_{2n+1} \left(\tilde{\eta} \right) \int 
  S_{2n+1} ( \tilde{r}( \tilde{\eta} ) ) 
  Q_{2n+1} \left( \tilde{\eta} \right) d\tilde{\eta}
  \nonumber \\
 && 
  - Q_{2n+1} \left( \tilde{\eta} \right) \int
  S_{2m+1} ( \tilde{r} ( \tilde{\eta} )) 
  P_{2m+1} \left( \tilde{\eta} \right) d\tilde{\eta} ,
  \nonumber \\
\end{eqnarray}
where $C_{2n+1}$ and $D_{2n+1}$ are constants.
Two extra constants of integration arise from the two
integrals in Eq.~(\ref{eq:g-s1}).
However these constants can be absorbed
into $C_{2n+1}$ and $D_{2n+1}$.  
Thus we can neglect the constants of integration
from the integrals in Eq.~(\ref{eq:g-s1}).
For given $n$, two constants $C_{2n+1}$ and $D_{2n+1}$ 
should be determined with the two boundary conditions
\begin{eqnarray}
\label{eq:bcd1}
 \lim_{\tilde{\eta} \rightarrow 1} 
  \Theta_{2n+1} (\tilde{\eta} ) & = & \mbox{regular} , \\
\label{eq:bcd2}
 \lim_{\tilde{\eta} \rightarrow \infty} 
  \Theta_{2n+1} (\tilde{\eta} ) & = & 0 .
\end{eqnarray}
We discuss the behaviors of the righthand side
in Eq.~(\ref{eq:g-s1}) as 
$\tilde{\eta} \rightarrow 1 $ and $\tilde{\eta} \rightarrow \infty$.
$P_{2n+1} \left( \tilde{\eta} \right)$ diverges 
when $\tilde{\eta} \rightarrow \infty$, 
while $Q_{2n+1} \left( \tilde{\eta} \right)$ diverges 
when $\tilde{\eta} \rightarrow 1$. 
Thus taking account of the boundary condition, we derive 
\begin{eqnarray}
\label{eq:g-s2}
 \Theta_{2n+1} (\tilde{\eta} )& = & 
  P_{2n+1} \left(\tilde{\eta} \right) \int_{\infty}^{\tilde{\eta}} 
  S_{2n+1} (\tilde{r} ( \tilde{\eta} ) ) 
  Q_{2n+1} \left(\tilde{\eta} \right) d\tilde{\eta} 
  \nonumber \\
 && 
  - Q_{2n+1} \left(\tilde{\eta} \right) \int_{1}^{\tilde{\eta}}
  S_{2n+1} ( \tilde{r} (\tilde{\eta} )) 
  P_{2n+1} \left(\tilde{\eta} \right) d\tilde{\eta} .
  \nonumber \\
\end{eqnarray}
Therefore, we obtain the formal solution for the scalar field
\begin{equation}
\label{eq:theta1-exp}
 \vartheta^{(1)} \left( \tilde{r}, \mu \right)
 = \mbox{const}+\tilde{\alpha}
   \sum_{n=0}^{\infty} \Theta_{2n+1} (\tilde{r}) P_{2n+1} ( -\mu ) ,
\end{equation}
where the radial function $\Theta_{2n+1}$ is given by 
\begin{widetext}
\begin{eqnarray}
\label{eq:Th-solution}
 \Theta_{2n+1} ( \tilde{r} )& = & 
  P_{2n+1} \left(\frac{\tilde{r}-1}{\tilde{\beta}} \right) 
  \frac{1}{\tilde{\beta}}\int_{\infty}^{\tilde{r}} 
  S_{2n+1} ( \tilde{r} ) 
  Q_{2n+1} \left(\frac{\tilde{r}-1}{\tilde{\beta}} \right) d\tilde{r}
  - Q_{2n+1} \left(\frac{\tilde{r}-1}{\tilde{\beta}} \right) 
  \frac{1}{\tilde{\beta}} \int_{1+\tilde{\beta}}^{\tilde{r}}
  S_{2n+1} ( \tilde{r}) 
  P_{2n+1} \left(\frac{\tilde{r}-1}{\tilde{\beta}} \right) d\tilde{r} .
  \nonumber \\
 & = & 
  \frac{1}{\tilde{\beta}} 
   \sum_{k=1}^{2} \sum_{m=0}^{n} \sum_{l=0}^{m+k} 
    c\left( m,n,k,l\right) 
   \left[
    P_{2n+1} \left(\frac{\tilde{r}-1}{\tilde{\beta}} \right) 
    \int_{\infty}^{\tilde{r}} 
    I_{2m-1,l-2k} ( \tilde{r} ) 
    Q_{2n+1} \left(\frac{\tilde{r}-1}{\tilde{\beta}} \right) d\tilde{r}
   \right. \nonumber \\
 && \left. 
  - Q_{2n+1} \left(\frac{\tilde{r}-1}{\tilde{\beta}} \right) 
  \int_{1+\tilde{\beta}}^{\tilde{r}}
  I_{2m-1,l-2k} ( \tilde{r}) 
  P_{2n+1} \left(\frac{\tilde{r}-1}{\tilde{\beta}} \right) d\tilde{r} 
  \right] .
\end{eqnarray}
\end{widetext}
Here we define
\begin{equation}
 \tilde{\beta} := \sqrt{1-\tilde{a}^2}.
\end{equation}
Hereafter, we neglect the constant term in Eq.~(\ref{eq:theta1-exp}).
Explicit forms for the solution can be found by performing
the integration in Eq.~(\ref{eq:Th-solution}) 
analytically or numerically.

In particular, we can find the analytic form for
the leading-order solution $\Theta_{1}$ as
\begin{widetext}
\begin{eqnarray}
 \label{eq:th1-solution}
 \Theta_{1} & = &
  \frac{3\kappa\tilde{a}}{2} 
  \left[ \frac{1}{\tilde{\beta}^2} \left\{ 
    \frac{1 +2 \tilde{\beta} + 2\tilde{\beta}^2 }
    {\left( 1+ \tilde{\beta} \right)^2}
  - \frac{\left( \tilde{r}+1-2\tilde{\beta}^2 \right)
    \left( \tilde{r} -1 \right)}
   { \left( 1- \tilde{\beta}^2 \right)
    \left( \tilde{r}^2 +1- \tilde{\beta}^2\right)}
  - \frac{2 \left( \tilde{r}^2+1 \right)}
   {\left( \tilde{r}^2 +1 - \tilde{\beta}^2 \right)^2} \right\} 
   \right. \nonumber \\
 && \left. 
  + \left\{ \frac{1}{\left( 1- \tilde{\beta}^2 \right)^{\frac{3}{2}}}
   \left( \pi - 2 \arctan \frac{\tilde{r}}
   {\sqrt{1-\tilde{\beta}^2}} \right)
  + \frac{1}{\left( 1- \tilde{\beta}^2 \right)^{2}} 
   \log \frac{\left( \tilde{r}-1+\tilde\beta \right)^2}
   {\tilde{r}^2+ 1 - \tilde{\beta}^2 } 
   \right\} \left( \tilde{r} -1\right)
  \right] .
\end{eqnarray}
\end{widetext}
On the other hand, it seems difficult to find analytic forms 
for higher-order solution $\Theta_{2n+1}$
in terms of elementary functions. 
However, we can obtain numerical solutions for $\Theta_{2n+1}  (n\geq 1)$, 
which are provided in the next section.

\begin{figure}[t]
 \includegraphics[width=\linewidth]{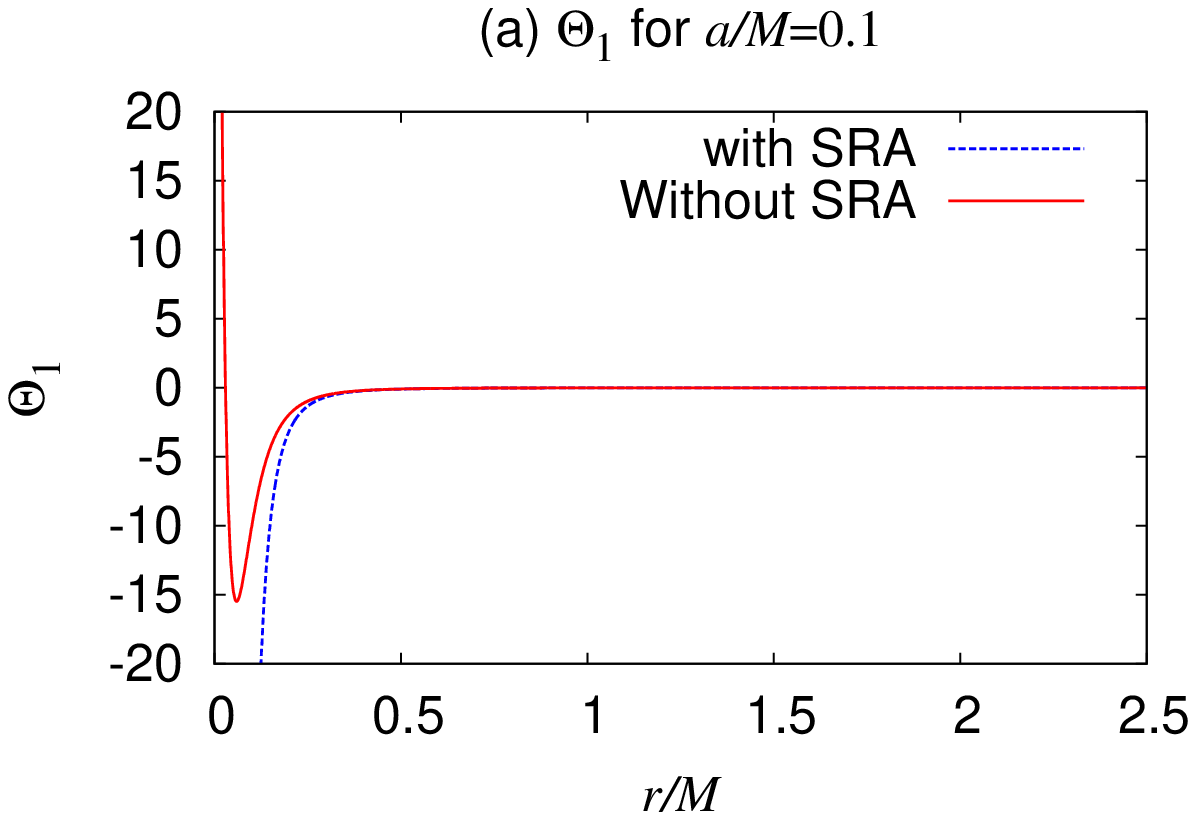}\\
 \includegraphics[width=\linewidth]{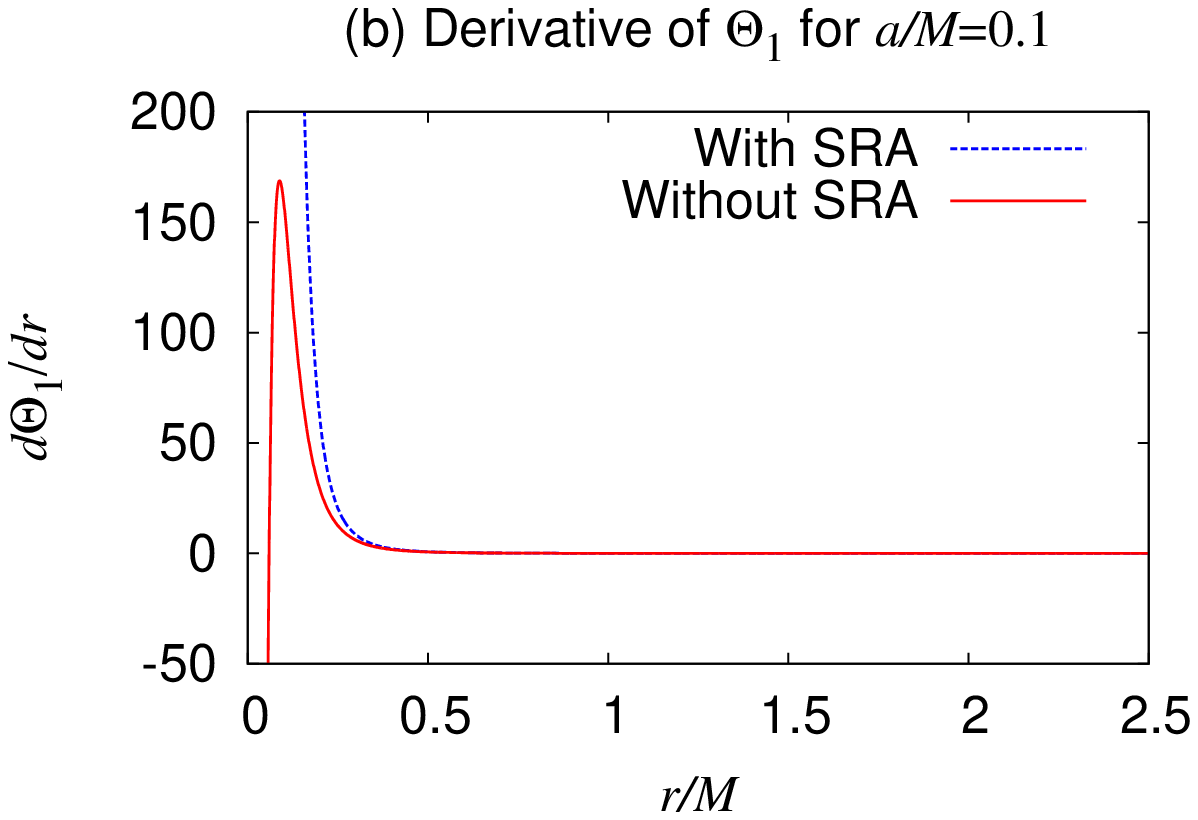}%
 \caption{\label{fig:Th1_a01}
  (a) Scalar function $\Theta_{1}$ and (b) its derivative 
  $d\Theta_{1}/d\tilde{r}$ with respect to $\tilde{r}$ when $\tilde{a}=0.1$. 
  The red curve shows our result, while the dashed blue curve shows 
  the result from the slow rotation approximation (SRA).
  The inner and outer horizons correspond to
  $\tilde{r}_{-} \simeq 0.005$ and $\tilde{r}_{+} \simeq 1.995$,
  respectively.}
\end{figure}
\begin{figure}[t]
 \includegraphics[width=\linewidth]{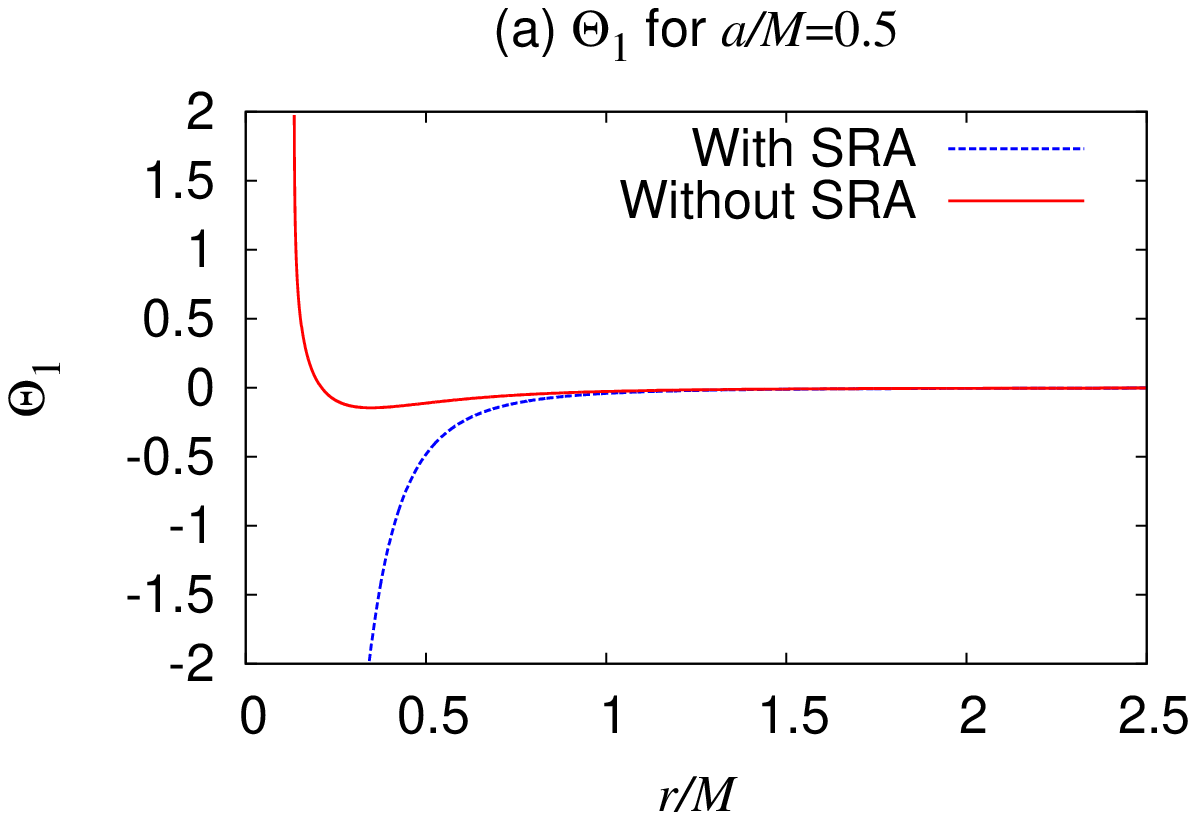}\\
 \includegraphics[width=\linewidth]{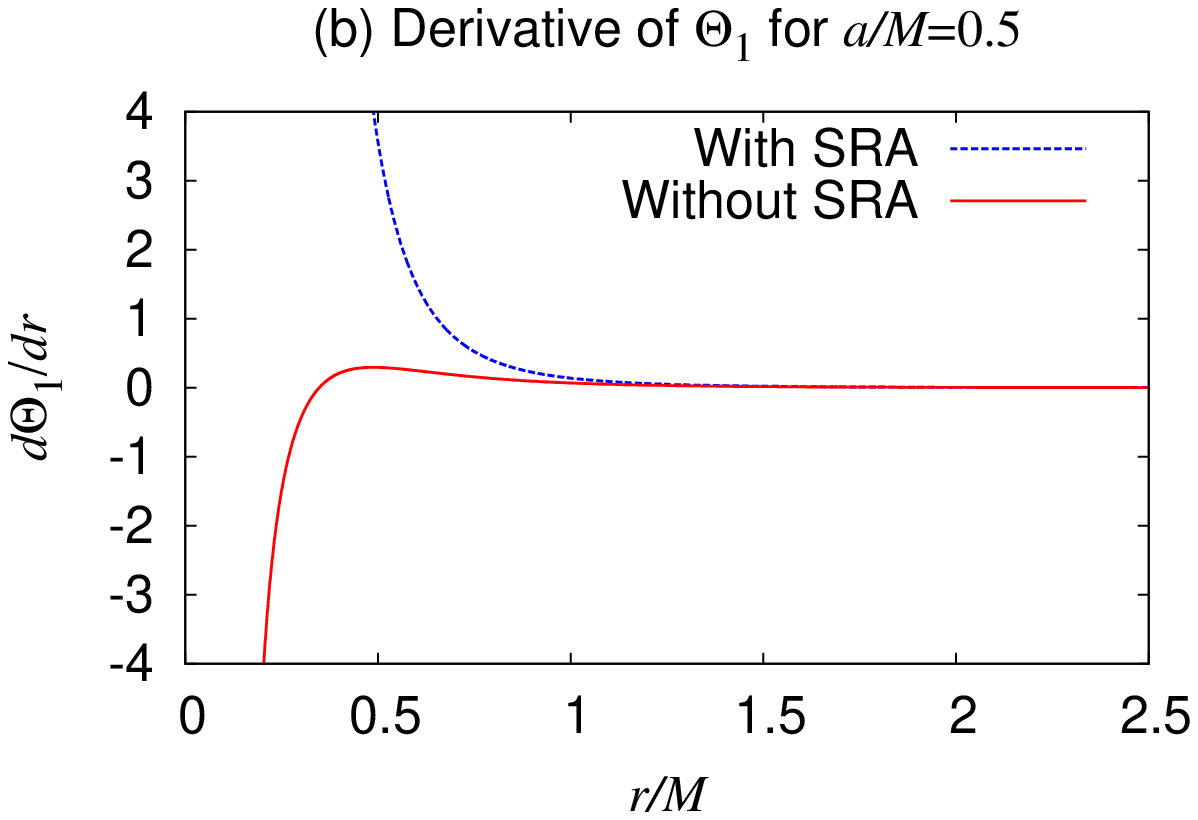}%
 \caption{\label{fig:Th1_a05}
   (a) Scalar function $\Theta_{1}$ and (b) its derivative 
  $d\Theta_{1}/d\tilde{r}$ with respect to $\tilde{r}$ when $\tilde{a}=0.5$.
  The red curve shows our result, while the dashed blue curve shows 
  the result from the slow rotation approximation (SRA).
  The inner and outer horizons correspond to
  $\tilde{r}_{-} \simeq 0.134$ and $\tilde{r}_{+} \simeq 1.866$,
  respectively.}
\end{figure}
\begin{figure}[t]
 \includegraphics[width=\linewidth]{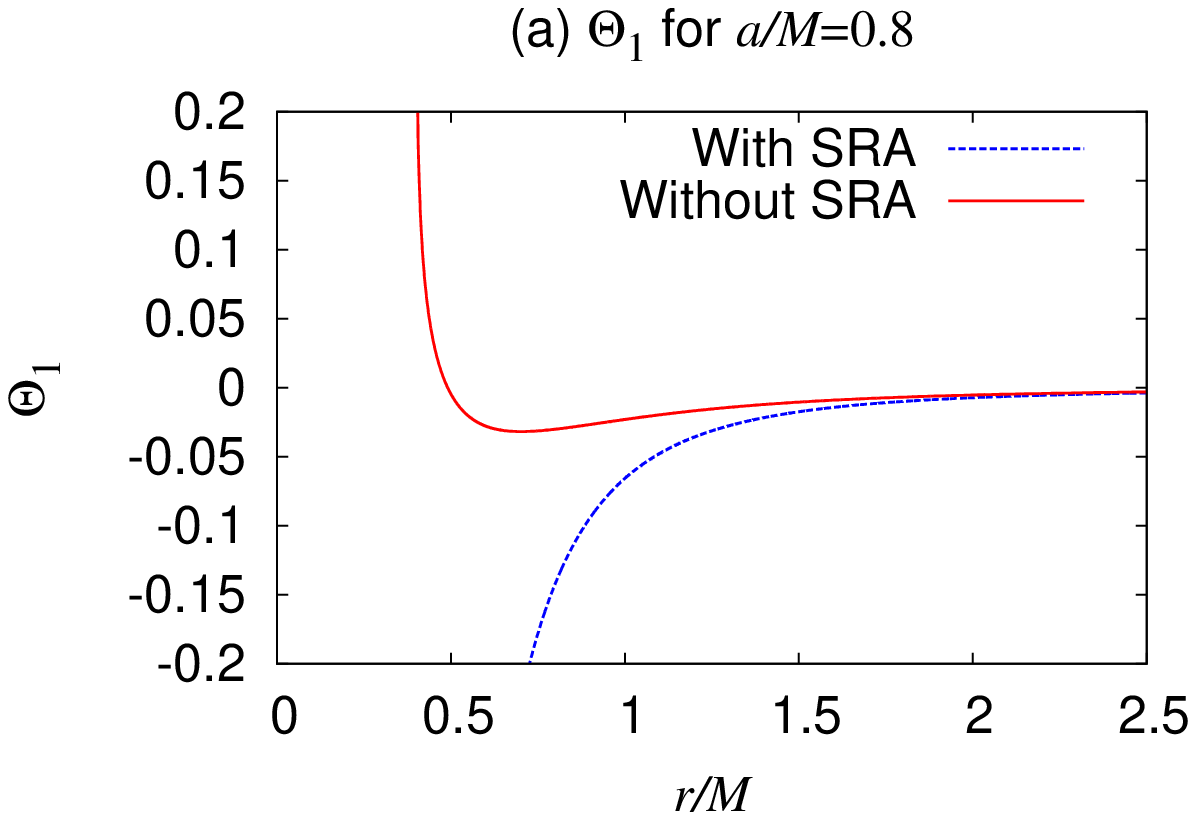}\\
 \includegraphics[width=\linewidth]{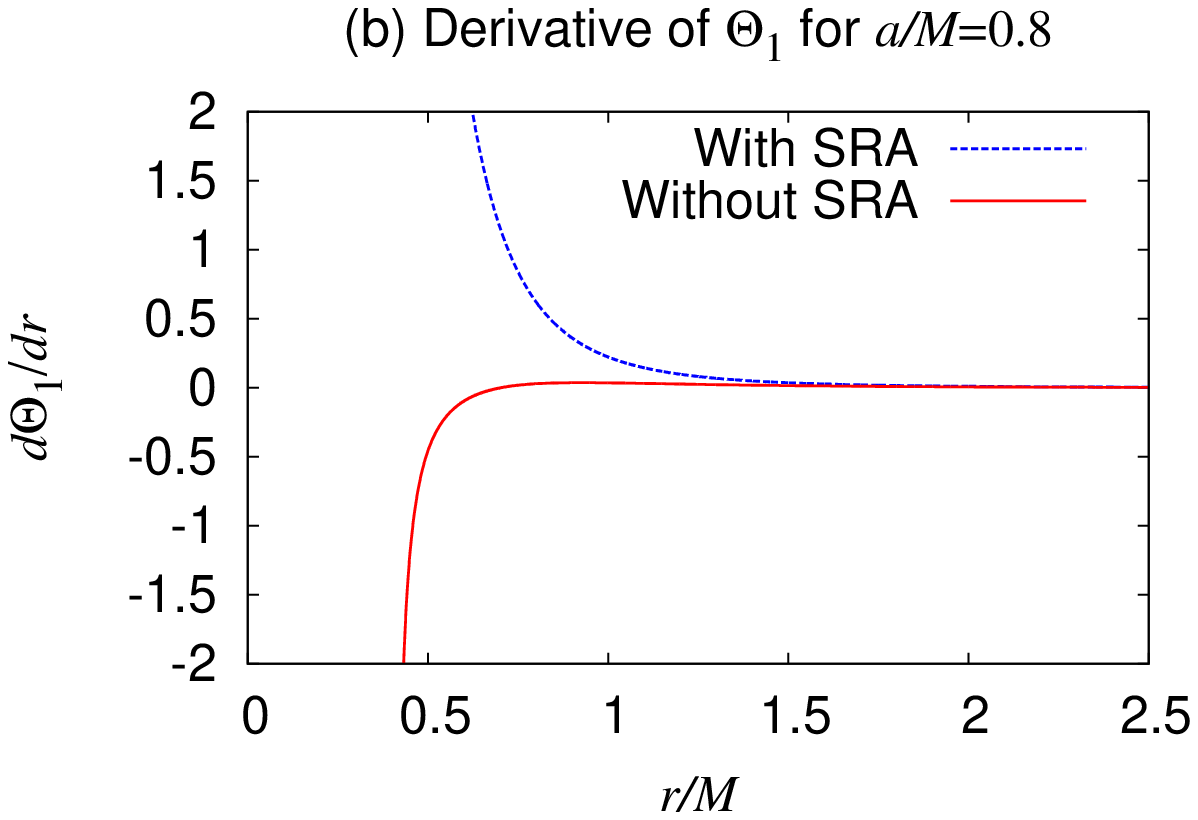}%
 \caption{\label{fig:Th1_a08}
   (a) Scalar function $\Theta_{1}$ and (b) its derivative 
  $d\Theta_{1}/d\tilde{r}$ with respect to $\tilde{r}$ when $\tilde{a}=0.8$.
  The red curve shows our result, while the dashed blue curve shows 
  the result from the slow rotation approximation (SRA).
  The inner and outer horizons correspond to
  $\tilde{r}_{-} \simeq 0.4$ and $\tilde{r}_{+} \simeq 1.6$,
  respectively.}
\end{figure}

\section{Properties of the scalar field solution}
\label{sec:prop-sf}

First of all, we discuss the properties of the leading-order 
solution $\Theta_{1}$ presented in the last section. 
Since the rotational parameter $\tilde{a}$ 
satisfies $0\leq \tilde{a} <1$, 
the power series expansion in $\tilde{a}$ is also valid
for the scalar field solution. As shown in Appendix 
\ref{sec:S_forms}, we roughly derive 
$S_{2n+1} \sim O\left( \tilde{a}^{2n+1} \right)$.
Thus, from Eq.~(\ref{eq:Th-diff-eq}), 
we conclude $\Theta_{2n+1} \sim O\left( \tilde{a}^{2n+1} \right)$.
Therefore, we recognize that
the leading-order solution $\Theta_{1}$ gives
the most dominant component in the 
first-order scalar field solution $\vartheta^{(1)}$, i.e., 
$\vartheta^{(1)} = \tilde{\alpha}\sum_{n}
 \Theta_{2n+1} P_{2n+1} \sim \tilde{\alpha} \Theta_{1} P_{1}$.
When we expand $\Theta_{1}$ into the power series 
in $\tilde{a}$, we derive
\begin{eqnarray}
\label{eq:th1-s-sr}
 \Theta_{1} & \simeq & 
   - \frac{\kappa\tilde{a}}{8 \tilde{r}^4} 
     \left( 5\tilde{r}^2 +10 \tilde{r} +18 \right)
   + \frac{\kappa\tilde{a}^3}{80 \tilde{r}^6} 
     \big( 5\tilde{r}^4 +10 \tilde{r}^3 +48 \tilde{r}^2
 \nonumber \\
 &&
     +152 \tilde{r} +400 \big) 
   + O\left( \tilde{a}^5 \right) .
\end{eqnarray}
Hence, from the first term of the righthand side 
of Eq.~(\ref{eq:th1-s-sr}), we retrieve the result 
of the slow rotation approximation \cite{yp, kmt2}
\begin{eqnarray}
 \vartheta^{(1)} \left( \tilde{r} , \theta \right)
 & \simeq & - \frac{\tilde{\alpha} \kappa\tilde{a}}{8 \tilde{r}^4} 
     \left( 5\tilde{r}^2 +10 \tilde{r} +18 \right) \cos \theta 
     + O\left( \tilde{a}^3 \right). \quad 
\end{eqnarray}
 From Eq.~(\ref{eq:th1-solution}), 
we also derive the asymptotic form of $\Theta_{1}$ for large $r$,  
\begin{eqnarray}
 \Theta_{1} & \simeq &
  - \frac{\kappa\tilde{a}}{2} 
    \frac{1 +2 \tilde{\beta} + 2\tilde{\beta}^2 }
    {\left( 1+\tilde{\beta}\right)^2} \frac{1}{\tilde{r}^2}
  - \kappa\tilde{a}
    \frac{1 +2 \tilde{\beta} + 2\tilde{\beta}^2 }
    {\left( 1+\tilde{\beta}\right)^2} \frac{1}{\tilde{r}^3} 
  \nonumber \\
 && 
  - \frac{3 \kappa\tilde{a}}{10} 
    \frac{5 +10 \tilde{\beta} + 11\tilde{\beta}^2 
     +2 \tilde{\beta}^3 + 2\tilde{\beta}^4}
    {\left( 1+\tilde{\beta}\right)^2} \frac{1}{\tilde{r}^4}
  \nonumber \\
 && + O\left( \tilde{r}^{-5}\right)  .
\end{eqnarray}
Thus $\Theta_{1}$ decays in proportional to 
$1/ \tilde{r}^{2}$ at infinity. At the outer 
horizon $\tilde{r}=\tilde{r}_{+}$, we derive
\begin{eqnarray}
 \Theta_{1} \left( \tilde{r}_{+} \right) 
 & = & \frac{3\kappa\tilde{\beta}}{4\tilde{a}^3}
  \Bigg[ -1 + \tilde{\beta} + 2\tilde{a}
  \left( \pi - 2 \arctan \frac{1+\tilde{\beta}}{\tilde{a}}\right)
  \nonumber \\
 && + 2 \log \frac{2\tilde{\beta}^2}{1+\tilde{\beta}}
  \Bigg] .
\end{eqnarray}
Thus $\Theta_{1}$ is finite at the outer horizon. 
However, it should be emphasized that 
$\Theta_{1}$ diverges when $\tilde{r}$ approaches  
the inner horizon $\tilde{r}=\tilde{r}_{-}:=1-\tilde{\beta}$. 
Near the inner horizon, $\Theta_{1}$ behaves as
\begin{eqnarray}
\label{eq:blowup_r-}
 \Theta_{1} \propto -\frac{3\kappa \tilde{\beta}}
   {\tilde{a}^3} \log \left| \tilde{r} -\tilde{r}_{-}\right| .
\end{eqnarray}
Therefore, $\Theta_{1}$ that satisfies the boundary conditions
(\ref{eq:bcd1}) and (\ref{eq:bcd2})
becomes problematic near the inner horizon, 
rather than the curvature singularity.

Figures \ref{fig:Th1_a01}, \ref{fig:Th1_a05} 
and \ref{fig:Th1_a08} show the function 
$\Theta_{1}$ and its derivative
$d\Theta_{1} / d\tilde{r}$ for $\tilde{a}=0.1$, 
$\tilde{a}=0.5$ and $\tilde{a}=0.8$.
We confirm that both $\Theta_{1}$ and 
$d\Theta_{1} / d\tilde{r}$ decay when $\tilde{r}$ increases
and are regular at the outer horizon $\tilde{r} = \tilde{r}_{+}$,
and that both diverge when $\tilde{r}$ approaches the 
inner horizon $\tilde{r} = \tilde{r}_{-}$.

\begin{figure}[t]
 \includegraphics[width=\linewidth]{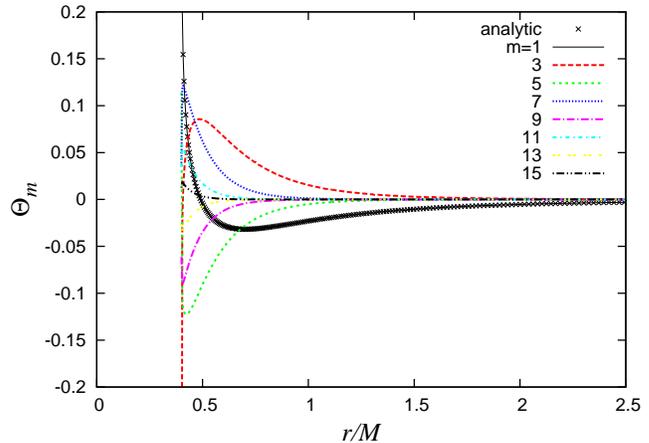}
 \caption{\label{fig:Thm_a08}
  $\Theta_{m} (\tilde{r})$ $(m=1,3,5,\cdots ,15 )$
  obtained by numerical integration when $\tilde{a}=0.8$.
  The result of $\Theta_{1}$ (solid curve)
  coincides with the analytic result (cross). 
  The inner and outer horizons correspond to
  $\tilde{r}_{-} \simeq 0.4$ and $\tilde{r}_{+} \simeq 1.6$,
  respectively.}
\end{figure}

Next, we discuss the properties of the higher-order 
functions $\Theta_m (m=1, 3, 5, 7, \cdots)$. 
For this purpose, we numerically calculate $\Theta_m$ 
based on the first line of Eq.~(\ref{eq:Th-solution}),
in which $\Theta_{2n+1}$ is expressed by the 
integrals containing $S_{2n+1}$, $P_{2n+1}$ and $Q_{2n+1}$. 
Here, $S_{2n+1}$ is given by Eq. (\ref{eq:smr}) with 
the source term $S(\tilde{r}, \mu)$ in Eq.~(\ref{eq:source-term}). 
The integrals are numerically calculated 
with the Romberg integration scheme 
in which Richardson extrapolations are applied repeatedly 
until desired numerical accuracy is achieved. 
The numerical code is made up based on the routine {\sf qromb} 
in \cite{nrecipe} with double precision. 
First, we numerically calculate $S_{2n+1}$. 
Then, we numerically calculate the integrals 
in the first line of of Eq.~(\ref{eq:Th-solution})
using the above method.
The higher-order terms of the Legendre polynomials 
of the first kind are calculated by using the recurrence 
relations between $P_n$, $P_{n+1}$ and $P_{n+2}$. 
The higher-order terms of the Legendre 
polynomials of the second kind are calculated 
from the definition by the hypergeometric functions 
and that by the relations between $P_n$ and $Q_n$ \cite{mathf}. 
Figure \ref{fig:Thm_a08} shows the numerical results 
of $\Theta_m$ for $\tilde{a}=0.8$ and $m=1, 3, 5, \cdots, 15$. 
The numerical result of $\Theta_1$ coincides with 
the analytic result. The lower-order functions $\Theta_n$ 
are basically larger than higher- order functions 
$\Theta_m (m>n)$ for the region of $\tilde{r}\gtrsim 1$. 
Each function is regular at the outer horizon. 
However, every functions $\Theta_m (m\ge 1)$ diverge 
when $\tilde{r}$ approaches $\tilde{r}_{-}$, 
i.e., the inner horizon. 
As in Eq.~(\ref{eq:blowup_r-}), the features of 
the divergence are different from those of power functions, 
but it is difficult to determine the functional forms 
from our numerical results at present.

\begin{figure}[t]
 \includegraphics[width=0.65\linewidth]{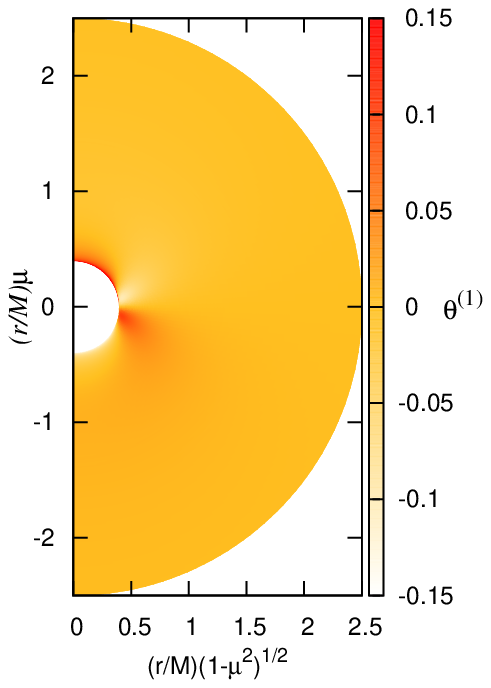}
 \includegraphics[width=\linewidth]{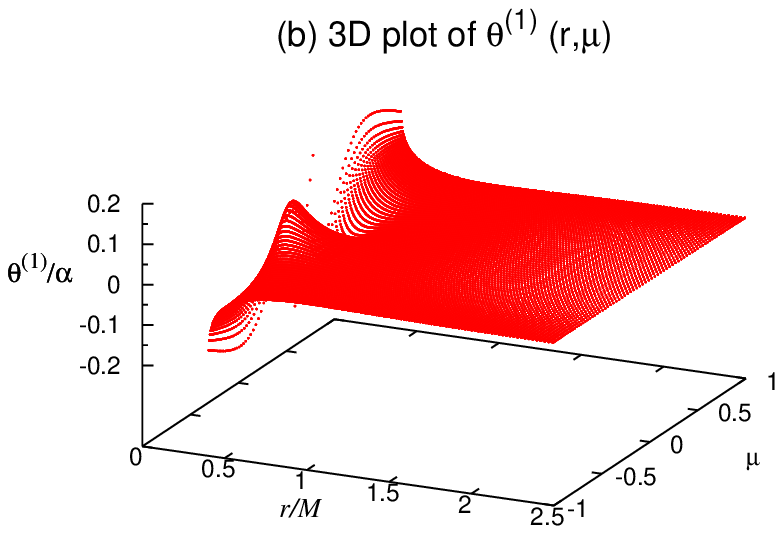}
 \caption{\label{fig:theta_a08}
  (a) Color map of $\vartheta^{(1)}$ for $\tilde{a}=0.8$. 
  The vertical axis denotes $\tilde{r} \mu$, and 
  the horizontal axis denotes $\tilde{r} \sqrt{1-\mu^2}$.
  (b) 3-dimensional plot of $\vartheta^{(1)} (\tilde{r}, \mu)$
  for $\tilde{a}=0.8$. The inner horizon 
  corresponds to $\tilde{r}_{-} \simeq 0.4$.}
\end{figure}

Next, we evaluate $\vartheta^{(1)}$ in Eq.~(\ref{eq:theta1-exp}) 
using the numerical results of $\Theta_m$. 
It should be noted that the series tends to 
converge when we raise the order to which we sum the series. 
Here, we define $\vartheta_{N}^{(1)}$
as the series summed up to the order $2N+1$, i.e.
$\ds\vartheta_{N}^{(1)} := \tilde{\alpha}
   \sum_{n=0}^{N} \Theta_{2n+1} P_{2n+1}$.
A relative error is defined as 
$\left( \vartheta_{N}^{(1)} - \vartheta_{N-1}^{(1)}\right)
   / \vartheta_{N-1}^{(1)}$. 
We confirm that when the series is truncated at $m=25$, 
the values of $\vartheta^{(1)}$ numerically 
converges within the relative error $\sim$1\%, 
except the divergent region on the inner horizon. 
Figure \ref{fig:theta_a08} shows (a) a color map 
of $\vartheta^{(1)}$ and (b) 3-dimensional plot 
of $\vartheta^{(1)}(\tilde{r}, \mu)$ for $\tilde{a}=0.8$. 
The values of $\vartheta^{(1)}$ are normalized 
by the coupling constant $\tilde{\alpha}$. 
We also confirm the signature that the scalar 
field solution $\vartheta^{(1)}$ diverges 
when $\tilde{r}$ approaches the inner horizon 
$\tilde{r}=\tilde{r}_{-}$.

Finally we discuss the divergence of the scalar field 
on the inner horizon again. When we take the slow 
rotation limit $\tilde{a}\rightarrow 0$, 
$\tilde{r}_{-}$ reduces to zero.  
Thus the location of the inner horizon coincides with the point 
of the curvature singularity in the slow rotation limit.
Therefore the problem about the divergence of the scalar field
has been hidden under the curvature singularity 
in the framework of the slow rotation approximation.
Without the slow rotation approximation, 
the scalar field becomes problematic at spacetime
points other than the curvature singularity.
The divergence of $\vartheta^{(1)}$ originates 
from that of $\Theta_{2n+1}$.
The radial function $\Theta_{2n+1}$ obeys 
Eq.~(\ref{eq:Th-diff-eq}), and the solutions for the 
homogeneous equation are given by the Legendre functions 
$P_{2n+1}(\tilde{\eta })$ and $Q_{2n+1} (\tilde{\eta} )$.
The function $Q_{2n+1} (\tilde{\eta} )$ 
diverges as $\tilde{\eta} \rightarrow \pm 1$, i.e.,
$\tilde{r} \rightarrow \tilde{r}_{\pm}$. 
Recalling Eq.~(\ref{eq:g-s2}), we understand that 
the divergence of $\Theta_{2n+1}$ at 
$\tilde{\eta} = 1$ can completely be
removed by adopting the appropriate domain of the integral.
However, the divergence of $\Theta_{2n+1}$ at 
$\tilde{\eta}=-1$ remains.
If we impose the boundary condition that
$\Theta_{2n+1}$ is regular at $\tilde{\eta}=-1$,
then $\Theta_{2n+1}$ would diverge at $\tilde{\eta}=1$.
Therefore, the divergence of the scalar field 
would inevitably occur on the inner horizon
when we impose its regularity on the outer horizon.

\section{Conclusion}
\label{sec:concl}

We have investigated the CS scalar field induced by a rapidly 
rotating black hole in the dynamical CS modified gravity. 
To discuss CS corrections, we adopted the perturbative 
approach assuming CS coupling to be weak. Furthermore, 
we took the Kerr spacetime to be the background to deal 
with rapid rotation of a black hole.
In this approach, the CS scalar field 
and the CS gravitational correction are determined alternately 
in the bootstrapping scheme based on the perturbation theory.
As the first step to the CS corrections, we obtained 
the solution for the scalar field analytically and numerically.
In the derivation, we assumed a stationary and axisymmetric 
configuration and adopted the boundary condition
that the scalar field is regular on the outer horizon 
of the Kerr black hole and vanishes at infinity.
In particular, we could find the analytic solution
for the leading term in the Legendre expansion of 
the scalar field with respect to the polar angle.
The scalar field is a well-behaved function outside 
the outer horizon.

However, the CS scalar field inevitably diverges on the inner 
horizon of the Kerr spacetime. Thus the scalar field becomes 
problematic when we approach the inner horizon. 
The readers might have a question: why should we care about 
the region inside the outer horizon?
If the scalar field had no effects on the spacetime,
the behavior of the scalar field inside the horizon
cannot be observed at all.
However, the scalar field, in fact, affects the structure 
of the spacetime via the stress-energy tensor. Note that 
the stress-energy tensor $T_{\vartheta}^{\mu\nu}$ is 
composed of the scalar field and its derivative. Therefore, 
the stress-energy tensor for the scalar field also diverges 
on the inner horizon. 
After all, ill-defined spacetime region increases in the 
dynamical CS modified gravity in comparison with general 
relativity.  The treatment for such a problematic region,
which might require quantum treatments, would be awaited 
in future works.

\begin{acknowledgments}
 We are grateful to Dr.~Leo C.~Stein for useful comments.

 {\it Note added in revision}.---After this paper was
 submitted, related works \cite{leo1,leo2} appear on the arXiv.
\end{acknowledgments}

\appendix
\section{Functional forms of $S_{2n+1}$}
\label{sec:S_forms}

We present the explicit forms of $S_{2n+1}$ 
derived from Eq.~(\ref{eq:s-term}) for lower orders.
\begin{enumerate}[(i)]
\item For $n=0$, we derive 
 \begin{equation}
  S_{1} 
   = \frac{3}{2\pi} \cdot
     \frac{3\tilde{a}\tilde{r} \left( \tilde{r}^2 -\tilde{a}^2 \right)}
     {\left( \tilde{r}^2 + \tilde{a}^2\right)^4} .
 \end{equation}
 The Taylor expansion of $S_{1}$ about $\tilde{a}=0$ gives 
 \begin{equation}
  S_{1} \simeq \frac{3}{2\pi} \left[ 
      \frac{3\tilde{a}}{\tilde{r}^5} 
    - \frac{15\tilde{a}^3}{\tilde{r}^7}
    + \frac{42\tilde{a}^5}{\tilde{r}^9} + O\left( \tilde{a}^7 
    \right) \right] .
 \end{equation}
\item For $n=1$, we derive 
 \begin{eqnarray}
  S_{3}
   & = & \frac{3}{2\pi} \cdot
     \frac{7}{12\tilde{a}^4 \left( \tilde{r}^2 + \tilde{a}^2\right)^4}
     \bigg[ - 
     ( 15 \tilde{a}\tilde{r}^7 + 55\tilde{a}^3 \tilde{r}^5
     + 73 \tilde{a}^5 \tilde{r}^3 
     \nonumber \\
  && + 57 \tilde{a}^7 \tilde{r})
     +15 \left( \tilde{r}^2 + \tilde{a}^2\right)^4 
     \arctan \frac{\tilde{a}}{\tilde{r}} \bigg] .
 \end{eqnarray}
 The Taylor expansion of $S_{3}$ about $\tilde{a}=0$ gives 
 \begin{equation}
  S_{3} \simeq \frac{3}{2\pi} \left[ 
    -\frac{10\tilde{a}^3}{\tilde{r}^7} 
    + \frac{392\tilde{a}^5}{9 \tilde{r}^9}
    - \frac{1260\tilde{a}^7}{11\tilde{r}^{11}} 
    + O\left( \tilde{a}^9 \right) \right] .
 \end{equation}
\item For $n=2$, we derive
 \begin{eqnarray}
  S_{5} 
   & = & \frac{3}{2\pi} \cdot
     \frac{11}{24\tilde{a}^6 \left( \tilde{r}^2 + \tilde{a}^2\right)^4}
     \bigg[ 
     1890\tilde{a}\tilde{r}^9 + 7035\tilde{a}^3 \tilde{r}^7
     \nonumber \\
  && + 9583 \tilde{a}^5 \tilde{r}^5 
     + 5533 \tilde{a}^7 \tilde{r}^3 + 1047 \tilde{a}^9 \tilde{r} 
     -105 \left( \tilde{r}^2 + \tilde{a}^2\right)^4 
     \nonumber \\
  && \times
     \left( 18\tilde{r}^2 + \tilde{a}^2\right) 
     \arctan \frac{\tilde{a}}{\tilde{r}} \bigg] .
 \end{eqnarray}
 The Taylor expansion of $S_{5}$ about $\tilde{a}=0$ gives 
 \begin{equation}
  S_{5} \simeq \frac{3}{2\pi} \left[ 
     \frac{112\tilde{a}^5}{9\tilde{r}^9} 
    - \frac{720\tilde{a}^7}{13 \tilde{r}^{11}}
    + \frac{1936\tilde{a}^9}{13\tilde{r}^{13}} 
    + O\left( \tilde{a}^{11} \right) \right] .
 \end{equation}
\item For $n=3$, we derive
 \begin{eqnarray}
  S_{7} 
   & = &  \frac{3}{2\pi} \cdot
     \frac{15}{32\tilde{a}^8 \left( \tilde{r}^2 + \tilde{a}^2\right)^4}
     \bigg[ - \big( 
     45045 \tilde{a} \tilde{r}^{11} + 179025\tilde{a}^3 \tilde{r}^9
     \nonumber \\
  && + 270354 \tilde{a}^5 \tilde{r}^7 + 188298 \tilde{a}^7 \tilde{r}^5 
     + 56665 \tilde{a}^9 \tilde{r}^3 
     \nonumber \\
  &&   +4805 \tilde{a}^{11} \tilde{r} \big)
     +315 \left( \tilde{r}^2 + \tilde{a}^2\right)^4 
     \nonumber \\
  && \times
     \left( 143\tilde{r}^4 + 44\tilde{a}^2 \tilde{r}^2  + \tilde{a}^4\right)
     \arctan \frac{\tilde{a}}{\tilde{r}} \bigg] .
 \end{eqnarray}
 The Taylor expansion of $S_{7}$ about $\tilde{a}=0$ gives 
 \begin{equation}
  S_{7} \simeq \frac{3}{2\pi} \left[ 
    -\frac{1440\tilde{a}^7}{143\tilde{r}^{11}} 
    + \frac{10560\tilde{a}^9}{221 \tilde{r}^{13}}
    - \frac{43680\tilde{a}^{11}}{323\tilde{r}^{15}} 
    + O\left( \tilde{a}^{13} \right) \right] .
 \end{equation}
\item For $n=4$, we derive
 \begin{eqnarray}
  S_{9} 
   & = & \frac{3}{2\pi} \cdot
     \frac{19}{384\tilde{a}^{10} \left( \tilde{r}^2 + \tilde{a}^2\right)^4}
     \bigg[ 2 \big( 
     3063060 \tilde{a}\tilde{r}^{13} 
     \nonumber \\
  && + 13258245 \tilde{a}^3 \tilde{r}^{11}
     + 22609587 \tilde{a}^5 \tilde{r}^9 + 18998298 \tilde{a}^7 \tilde{r}^7 
     \nonumber \\
  && + 7958786 \tilde{a}^9 \tilde{r}^5  
   +1445537\tilde{a}^{11}\tilde{r}^3 +70263 \tilde{a}^{13} \tilde{r}
   \big) 
     \nonumber \\
  && -6930 \left( \tilde{r}^2 + \tilde{a}^2\right)^4 
     \left( 884\tilde{r}^6 + 585\tilde{a}^2 \tilde{r}^4 
     +78\tilde{a}^4 \tilde{r}^2 + \tilde{a}^6 \right) 
     \nonumber \\
  && \times   \arctan \frac{\tilde{a}}{\tilde{r}} \bigg] .
 \end{eqnarray}
 The Taylor expansion of $S_{9}$ about $\tilde{a}=0$ gives 
 \begin{equation}
  S_{9} \simeq \frac{3}{2\pi} \left[
    \frac{1408\tilde{a}^9}{221\tilde{r}^{13}} 
    - \frac{1664\tilde{a}^{11}}{51 \tilde{r}^{15}}
    + \frac{38400\tilde{a}^{13}}{391\tilde{r}^{17}} 
    + O\left( \tilde{a}^{15} \right) \right] .
 \end{equation}
\item For $n=5$, we derive
 \begin{eqnarray}
  S_{11} 
   & = & \frac{3}{2\pi} \cdot
     \frac{23}{1536\tilde{a}^{12} \left( \tilde{r}^2 + \tilde{a}^2\right)^4}
     \bigg[ - \big( 
     218243025 \tilde{a}\tilde{r}^{15} 
     \nonumber \\
  && + 1033016985 \tilde{a}^3 \tilde{r}^{13} 
     + 1984607625 \tilde{a}^5 \tilde{r}^{11}
     \nonumber \\
  &&  
     + 1970934537 \tilde{a}^7 \tilde{r}^9 
     + 1062515883 \tilde{a}^9 \tilde{r}^7 
     \nonumber \\
  &&  
     + 296131667 \tilde{a}^{11} \tilde{r}^5 
     + 35848187 \tilde{a}^{13} \tilde{r}^3 
     \nonumber \\
  &&  +1134603 \tilde{a}^{15} \tilde{r}
   \big) +45045 \left( \tilde{r}^2 + \tilde{a}^2\right)^4 
     \nonumber \\
  && \times
     \big( 4845\tilde{r}^8 + 5168\tilde{a}^2 \tilde{r}^6  
     + 1530\tilde{a}^4 \tilde{r}^4 
     + 120 \tilde{a}^6 \tilde{r}^2 
     \nonumber \\
  && +\tilde{a}^8 \big)
     \arctan \frac{\tilde{a}}{\tilde{r}} \bigg] .
 \end{eqnarray}
 The Taylor expansion of $S_{11}$ about $\tilde{a}=0$ gives 
 \begin{equation}
  S_{11} \simeq \frac{3}{2\pi} \left[
    - \frac{3328\tilde{a}^{11}}{969\tilde{r}^{15}} 
    + \frac{6144\tilde{a}^{13}}{323 \tilde{r}^{17}}
    - \frac{17408 \tilde{a}^{15}}{285 \tilde{r}^{19}} 
    + O\left( \tilde{a}^{17} \right) \right] .
 \end{equation}
\end{enumerate}
Higher-order functions can also be obtained similarly.

\end{document}